\documentclass{article}
\usepackage{lingmacros}
\usepackage[colorlinks=true,linkcolor=blue,citecolor = blue]{hyperref}%
\usepackage{tree-dvips}
\usepackage{graphicx}
\usepackage{caption}
\usepackage{float}
\usepackage{amsmath}
\usepackage{amssymb}
\usepackage{amsthm}

\usepackage{tikz}
\usetikzlibrary{shapes,positioning}
\usepackage{mathtools}
\usepackage[super]{nth}
\usepackage{array}
\usepackage{tabularx}
\usepackage{lingmacros}
\usepackage{tree-dvips}
\usepackage{graphicx}
\usepackage{algorithm}
\usepackage{algpseudocode}
\usepackage{multirow}
\usepackage{makecell}
\usepackage{array}
\usepackage{titlesec}
\usepackage{natbib}
\usepackage{graphicx} 
\usepackage{authblk}

\title{PolyModel for Hedge Funds' Portfolio Construction Using Machine Learning}

\author[1]{Siqiao Zhao\thanks{These authors contributed equally to this work.}}
\author[2]{Dan Wang\thanks{These authors contributed equally to this work. Corresponding author email: \texttt{dwang35@stevens.edu} }}
\author[3]{Raphael Douady}
\affil[1]{Independent researcher}
\affil[2]{School of Business, Stevens Institute of Technology}
\affil[3]{Paris 1 Panthéon-Sorbonne University}
\date{}

\begin{document}

\maketitle

\begin{abstract}



The domain of hedge fund investments is undergoing significant transformation, influenced by the rapid expansion of data availability and the advancement of analytical technologies. This study explores the enhancement of hedge fund investment performance through the integration of machine learning techniques, the application of PolyModel feature selection, and the analysis of fund size. We address three critical questions: (1) the effect of machine learning on trading performance, (2) the role of PolyModel feature selection in fund selection and performance, and (3) the comparative reliability of larger versus smaller funds.

Our findings offer compelling insights. We observe that while machine learning techniques enhance cumulative returns, they also increase annual volatility, indicating variability in performance. PolyModel feature selection proves to be a robust strategy, with approaches that utilize a comprehensive set of features for fund selection outperforming more selective methodologies. Notably, Long-Term Stability (LTS) effectively manages portfolio volatility while delivering favorable returns. Contrary to popular belief, our results suggest that larger funds do not consistently yield better investment outcomes, challenging the assumption of their inherent reliability.

This research highlights the transformative impact of data-driven approaches in the hedge fund investment arena and provides valuable implications for investors and asset managers. By leveraging machine learning and PolyModel feature selection, investors can enhance portfolio optimization and reassess the dependability of larger funds, leading to more informed investment strategies.
\end{abstract}

\section{Introduction}
\subsection{Background}

\indent

Factor models are fundamental tools in finance, particularly in portfolio optimization and empirical asset pricing. These models hypothesize that the returns of securities or portfolios are influenced by a combination of common factors and unique idiosyncratic elements. Commonly, factor models are estimated using statistical methods such as principal component analysis (PCA)\citep{pca}, maximum likelihood estimation, and factor analysis. Within this framework, factor models are broadly categorized into single-factor and multifactor models.

The single-factor model, which typically uses market return as the sole factor, is valued for its simplicity and ease of interpretation. This simplicity facilitates straightforward implementation, making it practical for various applications. However, the single-factor model is limited by its explanatory power and does not account for diversification benefits. It assumes a linear relationship between asset returns $r_i$ and the market factor, which may not hold during periods of market volatility or regime shifts, where returns can display non-linear behaviors.

In contrast, the multifactor model incorporates multiple factors, such as interest rates and industry indices, allowing it to capture the complex dynamics and diversification of the financial markets more effectively. However, this model faces challenges such as overfitting, multicollinearity, and the difficulty of adapting to dynamic market conditions.

\subsection{Challenges on Hedge Funds' Data}

When employing multifactor models to analyze hedge fund returns, several issues emerge, as highlighted by researchers \citep{1b27312e-6c06-3abe-b265-3f6b6c3166fc} and \citep{doi:10.1146/annurev-financial-110311-101741}. The discussion by \citep{multi-poly-model} also aligns with our observations regarding the limitations of multifactor models in this context.

The application of multifactor models to hedge funds frequently encounters challenges such as overfitting, issues of correlation versus causation, nonlinearity, and nonstationarity. The inclusion of multiple factors increases the susceptibility to overfitting, where models may fit the training data well but fail to generalize to new data. Additionally, while a high correlation between factors and returns might be observed, this does not necessarily imply a causal relationship. Although multifactor models can provide substantial explanatory power for hedge fund returns, they do not guarantee that the identified factors are the primary drivers of a hedge fund's strategy.

Hedge fund returns are notably nonlinear due to the dynamic nature of portfolio strategies and the use of complex investment derivatives. Linear multifactor models struggle to capture critical information in the tails of hedge fund return distributions, where significant risks and opportunities often reside. Moreover, the performance of a hedge fund is influenced by various factors, including the skills of the fund manager, the chosen trading instruments, and the overarching investment strategy. Past performance explained by a multifactor model does not necessarily predict future results.

To address these challenges, we propose the PolyModel theory, which is discussed in detail in Section 2.1. This theory offers a more flexible and robust framework for analyzing hedge fund returns, accommodating the complex dynamics and inherent nonlinearity of this investment domain.

\subsection{Addressing the Challenges: The PolyModel and Machine Learning Fusion}

The integration of the PolyModel theory with advanced machine learning algorithms provides a sophisticated and precise framework for understanding hedge fund returns. This combination addresses the inherent challenges in hedge fund data analysis, offering a robust methodology that enhances both the accuracy and reliability of investment decisions.

Machine learning is renowned for its capacity to tackle complex problems across various fields, including finance \citep{wang2021application, wang2023alerta, chen2020bridging, Xu_2024}, recommendation systems \cite{dong2024musechat}, social media sentiment analysis \cite{zhu2021twitter}, transportation \cite{hu2023artificial}, cloud computing \cite{ali2025enabling, ma2024malletrain} etc. Its core strength lies in its ability to autonomously learn from data, recognize patterns, and make informed predictions or decisions. This capability allows machine learning algorithms to continuously improve as they process more data, making them exceptionally effective for managing large-scale, high-dimensional, and nonlinear datasets. In finance, machine learning has proven effective in numerous applications such as sentiment analysis, portfolio optimization, risk management, and the development of trading strategies \citep{golbayani2020application, wang2023sparsity, zhu2021twitter}. Leveraging these capabilities enables the creation of sophisticated models that more accurately reflect the complex dynamics of financial markets, thereby enhancing investment outcomes.

For example, \citet{vijh2020stock} demonstrated that machine learning techniques could outperform traditional linear models in stock price prediction. \citet{wang2020image} investigated the use of convolutional neural networks (CNNs) for analyzing financial data from the EDGAR database, discovering that imaging techniques were particularly effective for analyzing financial ratio data, though they did not significantly enhance the analysis of fundamental data. Additionally, \citet{ma2021portfolio} successfully applied machine learning to optimize portfolio selection, achieving notable improvements in risk-adjusted returns.

Conversely, the PolyModel theory offers a flexible and comprehensive approach to modeling hedge fund returns. It incorporates a diverse array of risk factors and accommodates nonlinear relationships between assets and factors, effectively addressing the limitations commonly associated with traditional multifactor models, such as overfitting, multicollinearity, and the challenge of adapting to dynamic market conditions.

The fusion of the PolyModel theory with cutting-edge machine learning algorithms not only enriches our understanding of hedge fund returns but also enhances the methodology for analyzing these returns. This innovative approach overcomes significant challenges in hedge fund data analysis, leading to more informed and effective investment strategies.

\subsection{Contributions of This Paper}

This paper pioneers a groundbreaking approach that integrates financial theory with state-of-the-art machine learning techniques, representing a significant leap forward in the analysis of hedge funds. The principal contributions of this study are outlined as follows:

\begin{itemize} 
\item \textbf{Integration of PolyModel and Machine Learning:} We introduce a sophisticated framework that combines PolyModel theory with advanced machine learning algorithms. This integration enhances the precision and depth of our understanding of hedge fund returns, offering insights that are more nuanced and accurate than previously possible. 
\item \textbf{Enhanced Analytical Methodology:} Our approach addresses and overcomes the limitations of traditional multifactor models by tackling issues such as nonlinearity, nonstationarity, and the dynamic nature of hedge fund strategies. This results in a more robust and adaptable methodology for hedge fund data analysis. 
\item \textbf{Advanced Insights into Feature Selection:} We explore the significant impact of feature selection through PolyModel on trading performance. This investigation provides critical guidance for developing effective fund selection strategies that are empirically grounded. 
\item \textbf{Reevaluation of Fund Reliability Metrics:} Contrary to the conventional wisdom that larger funds are more reliable, our findings challenge this notion and illuminate new perspectives on fund allocation based on Assets Under Management (AUM). This reevaluation prompts a reconsideration of how reliability is assessed in the context of hedge fund investments. 
\end{itemize}

The structure of the paper is as follows: Section 2 delineates the methodology that merges PolyModel with machine learning for a comprehensive analysis of hedge funds. Section 3 discusses the empirical results, shedding light on how feature selection and fund size influence trading performance. Section 4 concludes the study, summarizing key findings and implications.

\section{Methodology}
\subsection{PloyModel Theory}\label{polymodel}

In the introduction, we already have a review of the traditional single and multi factor models, and their applications in finance. Moreover,
we analyzed why they may not be the best choices for financial portfolio
constructions, especially for the cases where asset time series data is sparse,
making it challenging to apply conventional statistical or machine learning
techniques. Such scenarios are common in the finance industry, especially
concerning the performance of hedge funds, where returns and AUMs are
only reported on a monthly basis, and historical data is limited.

To overcome these chanllenges, \cite{org-polymodel} proposed a framework called PolyModel theory. Breifly speaking, PolyModel theory aims to study the reactions of target time series with respect to the evolution of each risk factors siting inside a large factor pool such that these factors can be regarded as a very good proxy of our real world, not limited to finance only. 

By finding quantitative measures to characterize these reactions and combine them in a creative way, one can better understand the many faces of the target time series in a unified way. \cite{org-polymodel} also lied the theoretical foundations of the PolyModel theory and proved that under mild assumptions of the financial and risk factor time series, PolyModel barely loses any information by looking at each single risk factor first and then combining the results together. More detials and its application can be found in  \cite{org-polymodel}, \cite{sq-thesis} and \cite{202409.1453}.

In this paper, our main focus is the application of PolyModel theory and boost its power with machine learning techniques. Let's start from the practical implementation of PolyModel.

\indent

 To apply the PolyModel Theory, let's first define two fundamental components which are the fundamental building blocks:
 
 \begin{itemize}
     \item A pool of target assets $\{Y_i\}_{i \in I}$: these are the assets which will be used for the portfolio constructions.
     \item A pool of risk factors $\{X_j\}_{j \in J}$: each of the factors is the independent variable in one simple linear regression; the main requirement for the pool of risk factors is that this pool needs to cover various aspects of the financial market regardless the the sectors of the target assets, moreover, it is the best if the time series of these factors have a long enough history so that different historical scenarios and rare events can be captured.
 \end{itemize}

\subsubsection{Formulation of PolyModel in Practice}\label{model}

\indent

 For every target time series $Y_i$, $i \in I$, there is a collection of simple linear regression models (which can contain nonlinear terms of independent variables):
 \begin{equation}
     \begin{cases}
        Y_i = \Phi_{i1}(X_1) + \epsilon_1\\
        Y_i = \Phi_{i2}(X_2) + \epsilon_2 \\
        ... ... \\
        Y_i = \Phi_{in}(X_n) + \epsilon_n
    \end{cases}
 \end{equation}
where
\begin{itemize}
    \item $n$ is the number of the risk factors, in other words, $n = \# J$.
    \item $\Phi_{ij}$ is a polynomial which is assumed to capture the major relationship between independent variable $X_j$ and dependent variable $Y_i$.
    \item $\epsilon_{j}$ is the noise term in the regression model with zero mean; usually it is assumed to be normal distribution but does not have.
\end{itemize}

As shown in (1), for each pair $Y_i$ and $X_j$, we need to fit a simple regression model between them $Y_i = \Phi_{ij}(X_j) + \epsilon_j$. $\Phi_{ij}(x)$ is chosen to be a polynomial of degree 4; the reason is that a too large degree will introduce overfitting while a degree 4 polynomial can represent increasing, decreasing, convex, concave curves or even curves with inflection points. Thus, a dgree 4 polynomial is intuitively a good choice between the trade off flexibility and overfitting.

Moreover, we further assume that
\begin{equation}
    \Phi_{ij}(x) = \Sigma_{k = 0}^{4} \beta_{ij}^k H_{k}(x),
\end{equation}
where $H_{k}(x)$ is the Hermitian polynomial of degree $k$.

For each target and risk factor pair $(Y_i, X_j)$, assume that we are given their observations: $Y_i$ and $X_j$ for time $t = 1,2, ..., T$, to estimate the coefficients $\beta_{ij}^k$ using the techniques from ordinary least squares method and its generalizations, let's first write the regression into matrix form

\begin{equation}
    \overrightarrow{Y_i} = \bold{H}_j^{T} \overrightarrow{\beta_{ij}} + \overrightarrow{\epsilon_{ij}},
\end{equation}

\subsubsection{Estimation of PolyModel}

\indent

From (3), let's consider the square sum of errors in the matrix format, and the goal is to minimize it, and the estimated coefficients are
\begin{equation}
    \widehat{\overrightarrow{\beta_{ij}}} := arg \  min_{\{\overrightarrow{\beta_{ij}} \in R^5\}} 
    (\overrightarrow{Y_i} - \bold{H}_j^{T} \overrightarrow{\beta_{ij}})^T (\overrightarrow{Y_i} - \bold{H}_j^{T} \overrightarrow{\beta_{ij}}).
\end{equation}

To get a closed-form solution for $\widehat{\overrightarrow{\beta_{ij}}}$, we set
\begin{equation*}
    \frac{\partial (\overrightarrow{Y_i} - \bold{H}_j^{T} \overrightarrow{\beta_{ij}})^T (\overrightarrow{Y_i} - \bold{H}_j^{T} \overrightarrow{\beta_{ij}})}{\partial \overrightarrow{\beta_{ij}}} = -2 \boldsymbol{H}_j (\overrightarrow{Y_i} - \bold{H}_j^{T} \overrightarrow{\beta_{ij}}) = 0
\end{equation*}
which (at least formally) implies that 
\begin{equation}
    \widehat{\overrightarrow{\beta_{ij}}} = (\bold{H}_j \bold{H}_j^{T})^{-1} \bold{H}_j \overrightarrow{Y_i}.
\end{equation}
One potential issue about formula (5) is that square matrix $\bold{H}_j \bold{H}_j^{T}$ is not guaranteed to be invertible while generalized matrix inverse may introduce more uncertainty or inaccuracy.

Now the benefits of using Hermitian polynomial come into the picture. It is one of the Hermitian polynomials' signature properties that they form an orthogonal basis with respect to standard normal density function (see the proof in Appendix). Thus, $E[\bold{H}_j \bold{H}_j^{T}] = I_{5 \times 5}$ when the $X_i(t)'s$ are i.i.d, and follow standard normal distribution. When $X_i$ represents the log return of some risk factor, this assumption can be approximated in a good sense, and helps to reduce the chance that $\bold{H}_j \bold{H}_j^{T}$ is not invertible.

However, there is still no guarantee that we can calculate the inverse of $\bold{H}_j \bold{H}_j^{T}$; one solution is to apply some regularity term in (4). In particular, we consider the popular ridge regularity with our original loss function (4), and get

\begin{equation}
    \widehat{\overrightarrow{\beta}}_{ij, \lambda} := arg \  min_{\{\overrightarrow{\beta_{ij}} \in R^5\}} 
    [(\overrightarrow{Y_i} - \bold{H}_j^{T} \overrightarrow{\beta_{ij}})^T (\overrightarrow{Y_i} - \bold{H}_j^{T} \overrightarrow{\beta_{ij}}) + \lambda ||\overrightarrow{\beta_{ij}}||^2,
\end{equation}
where
\begin{itemize}
    \item $||\overrightarrow{v}||^2$ denotes the $L_2$-norm of vector $\overrightarrow{v}$; assume that $\overrightarrow{v} = (v_1, v_2, ..., v_n)^T$, $||\overrightarrow{v}||^2 = \Sigma_{i = 1}^{n} v_i^2$.
\end{itemize}

Ridge penalty can help to reduce the magnitudes of the coefficients, thus, can help to prevent overfitting. Moreover, ridge regression has closed-form solution. Actually, since the $L_2$-norm is a differentiable function of the vector entries, we can use the same method to derive the solution as how we get (6) out of (5), and the calibrated coefficients of ridge regression are

\begin{equation}
    \widehat{\overrightarrow{\beta}}_{ij, \lambda} = (\bold{H}_j \bold{H}_j^{T} + \lambda I_{5 \times 5})^{-1} \bold{H}_j \overrightarrow{Y_i},
\end{equation}

\indent

notice that the ridge penalty weight $\lambda$ makes the matrix $\bold{H}_j \bold{H}_j^{T}$ shrinkage to the diagonal matrix, making $\bold{H}_j \bold{H}_j^{T} + \lambda I_{5 \times 5}$ invertible.

\indent

We can see that now the fitted coefficients are functions of the hyper-parameter $\lambda$; to determine the optimal value for each simple regression, one can apply any state-of-art hyper-parameter tuning trick such as grid search plus cross-validation. However, we would like to point out that in PolyModel theory, we need to deal with a huge amount of risk factors, and our polynomial in the regression equation is only of degree 5, thus, our major concern for using ridge regression is to make the matrix $\bold{H}_j \bold{H}_j^{T} + \lambda I_{5 \times 5}$ invertible, thus, we usually choose a relatively small number as the value of $\lambda$ for all the target time series and risk factor pairs.

\subsubsection{Factor Importance and Selection: Target Shuffling and $P$-Value Score}\label{feature}

Recall that one of the goal of PolyModel theory is to find the set of risk factors which are most important to the target time series after fitting hundreds of simple regressions. How do we combine all the information and utilize the significant ones? This is the central topic of this section.

\indent

As the importance of different risk factors will evolve along with the time, we need to measure their significance at every time stamp. This reflects different target time series will have different reactions to the various risk factors during different regimes. Now let's discuss how we can identify the importance of features/risk factors in PolyModel. Let's look at the detials of target shuffling and $P$-value score.

\indent

To avoid fake strong relationship between target and risk factors, we apply target shuffling which is particular useful to identify cause-and-effect relationship. By shuffling the the targets, we have the chance to determine if the relationship fitted by the regression model is really causal by checking the probability of the $R^2$ we have seen based on the observations.

\indent

The procedure can be summarized as follows:
\begin{itemize}
    \item Do random shuffles on the target time series observations many times, say N times. For each $X_j$, let we assume that there are T data points $\{(Y_i(t_k),X_j(t_k)\}^T_{k=1}$. We fix the order of $X_j(t_k)$, and we do N times of random shuffle of $Y_i(t_k)$. In this way, we try to break any relation from the original data set and create any possible relations between the target and risk factor.
    
    \item For each newly ordered target observations $\{(Y'_i(t_k),X_j(t_k)\}^T_{k=1}$, we can fit a simple regression model and calculate the $R^2$. Then we get 
    
    \begin{center}
    $R^2_{shuffle} = \{R^2_{(1)},R^2_{(2)}, \cdots, R^2_{(N)}\}$.
    \end{center}
    
    Thus, we have a population of $R^2$ based on above procedures.
    
    \item Evaluate the significance of the $R^2$ calculated from the original data, for instance, we can calculate the p-value of it based on the $R^2$ population from last step. Here we assume that our original $R^2$ for target asset $Y_i$ and risk factor $X_j$ is denoted as $R^2_{ij}$. Then, we could define 
    
    \begin{center}
    $p_{ij} = P(R^2 > R^2_{ij})$.
    \end{center}
    
    \item We compute $-log(p_{ij})$ and call it $P$-Value Score of target asset $Y_i$ and risk factor $X_j$ which indicates the importance of the risk factor $X_j$ to the target asset time series $Y_i$.
    
\end{itemize}

\indent

The higher the $P$-Value Score is, the more important the risk factor is. As we  also need to take different regimes over the time into the picture, at each time stamp, we only look at the past 3 years' return data, and thus, we can have a dynamic $P$-Value Score series for each target asset $Y_i$ and risk factor $X_j$ pair.

\subsection{Statistical Features}

In this section, we will mainly introduce several statistical indicators
which will play the central roles in the hedge fund portfolio construction besides the very basic features such as the return and assets under management (AUM) of hedge funds.

\subsubsection{MRaR: Morningstar Risk-Adjusted Return}

\begin{center}
    $MRaR = (\frac{1}{T}\Sigma_{i=1}^{T}(1 + r_{Gt})^{-\gamma})^{-\frac{T}{\gamma}} - 1$,
\end{center}

\begin{center}
    $r_{Gt} = (\frac{1+r_t}{1+r_f}) - 1$,
\end{center}

where T is the total number of months in calculation period; $r_{Gt}$ is the geometric excess return at month t; $r_t$ is the asset return at month t; $r_f$ is the risk free return at month t; $\gamma$ is the risk aversion parameter, and Morningstar uses 2. We will adjust the value of $\gamma$ according to our risk flavor.

The Morningstar Risk-Adjusted Return (MRAR) serves as a metric to evaluate the performance of an investment or portfolio while taking the associated risk into consideration. The computation of MRAR involves several steps. Firstly, the total return of the funds for a specified time t is calculated.  Then, we need to set the benchmark term, and usually the benchmark is less risky. The calculation of the excess return is then conducted, involving the subtraction of the benchmark return from the total return at time t. This calculation aims to quantify the additional return achieved due to assuming additional risks beyond what could have been realized by adhering to a benchmark. Furthermore, MRAR is capable of quantifying the variability of the fund's performance within a specified time period T. A larger penalty effect is shown if the variation of the performance for the funds is significant.

\subsubsection{Sharpe Ratio}

Sharpe Ratio is a standard statistical measure indicating the performance of one's portfolio. It can be regarded as the return over the risk. Now, let's look at its formal definition.

\indent

Assume $R_{portfolio}$ represents the return of one's portfolio, $R_{b}$ represents the return of the benchmark financial time series, usually, it is the risk-free rates. Then Sharpe Ratio is defined as
\begin{center}
    Sharpe Ratio := $\frac{E(R_{portfolio} - R_{b})}{\sqrt{var(R_{portfolio} - R_{b})}}$.
\end{center}

\indent

Notice that when the benchmark is quite static (for instance, the risk-free rate), one may simplify Sharpe Ratio and replace it by $\frac{E(R_{portfolio} - R_{b})}{\sqrt{var(R_{portfolio})}}$ or even $\frac{E(R_{portfolio})}{\sqrt{var(R_{portfolio})}}$.

\subsubsection{SVaR}

\indent

SVaR can be regarded as a good alternative risk measure instead of VaR, in fact, it can be regarded as a factor model-based VaR. However, its strength resides in the modeling of nonlinearities and the capability to analyze a very large number of potential risk factors \cite{coste2009stressvar}. 

\indent

There are three major steps in the estimation of StressVaR of a hedge fund $Y_i$.

\begin{enumerate}
    \item Most relevant risk factors selection: for each risk factor $X_j$, we can calculate the $P$-Value Score of it with respect to $Y_i$. Recall Section 2.5.2, this score can indicate the explanation power of risk factor $X_j$, and the application of target shuffling improves the ability of our model in preventing discovering non-casual relations. Once a threshold of $P$-Value Score is set, we can claim that all the risk factors $X_j$ whose $P$-Value Score is above the threshold are the most relevant risk factors, and denote the whole set of them as $\Gamma_{i}$.
    
    \item Estimation of the Maximum Loss of $Y_i$:
    For every risk factor $X_j \in \Gamma_i$, using the fitted polynomial for the pair $(Y_i, X_j)$, we can predict the return of $Y_i$ for all risk factor returns from $1st$ to $99th$ quantiles of the risk factor distributions. In particular, we are interested in the potential loss of $Y_i$ corresponding to $\alpha \% = 98 \%$ of the factor returns. Once this is estimated for one factor $X_j$, we can define $SVaR_{i,j}$ for the pair $(Y_i, X_j)$ as follows:
    
    \begin{equation*}
        SVaR_{i,j} := \sqrt{\hat{Y}_{i,j, max}^2 + \sigma(Y_i)^2 \cdot (1 - R^2) \cdot \xi^2}
    \end{equation*}
    where
    \begin{itemize}
        \item $\hat{Y}_{i,j, max}$ is the maximum potential loss corresponding to $\alpha$ quantile of risk factor $X_j$.
        \item $\sigma(Y_i)^2 \cdot (1 - R^2)$ is unexplained variance under the ordinary least square setting which can be estimated by the following unbiased estimator if penalty terms are added to the regression models
        \begin{center}
            $\frac{\Sigma (Y_i - \hat{Y}_i)^2}{n-p}$,
        \end{center}
        where $p$ is the degree of freedom of the regression model.

        \item $\xi = \varphi^{-1}(\alpha) \approx 2.33$ where $\varphi $ is the cumulative distribution function (cdf) of standard normal distribution.

    \end{itemize}
    
    \item Calculation of StressVaR: The definition of StressVaR of $Y_i$ is
    \begin{center}
        $SVaR_i := max_{j \in \Gamma_i} SVaR_{ij} $.
    \end{center}
    
\end{enumerate}

\subsubsection{Long-term Alpha LTA}

\indent

For the given hedge fund and risk factor pair $(Y_i, X_j)$, assume we already fitted the regression polynomial $\Phi_{ij}(x)$. Assume that $\theta_{j, q}$ represents the $q$-quantile of the empirical distribution of $X_j$ where $q = 1\%, \  16\%, \  50\%, \  84\%, \  99\%$. They are calculated using the very long history of the factor. The extremes $1\%$ and $99\%$ are computed by fitting a Pareto distribution on the tails.

\indent

Then we define
\begin{center}
    $LTA(Y_i, X_j) := \Sigma_{q = 1 \%}^{99 \%} w_q \Phi_{ij}(\theta_{j, q})$,
\end{center}
subject to $E(X_j) = \Sigma_{q = 1 \%}^{99 \%} w_q \theta_{j, q}$, where $w_q$ correspond to Lagrange method of interpolating an integral and are hyper-parameters.

\indent

The global LTA (long-term average) is the median of the factor expectations for selected factors. $LTA_i$ for $Y_i$ is defined as the $50th$ quantile among all the LTA($Y_i$, $X_j$) values, where $X_j \in \Gamma_i$ represents the selected ones.

\subsubsection{Long-term Ratio LTR}

\indent

Once we get the $LTA_i$ and $SVaR_i$ for $Y_i$, $LTR_i$ is simply defined as

\begin{center}
    $LTR_i := \frac{LTA_i}{SVaR_i}$.
\end{center}

\subsubsection{Long-term LTS}

\indent

For fund $Y_i$, $LTS_i := LTA_i - \kappa \cdot SVaR_i$ where $\kappa$ is a hyper-parameter whose value is set to $5\%$.

\subsection{Portfolio Construction with Machine Learning}

In this section, we detail the methodology for constructing portfolios using machine learning techniques based on previously derived features.

For each hedge fund $Y_{i}$ at any given time, we consider five key features: Long-Term Stability ($LTS_{i}$), Maximum Return at Risk ($MRaR_{i}$), Sharpe Ratio, monthly return, and Assets Under Management (AUM). These features are utilized to train predictive models, such as XGBoost, to forecast the trend of returns for the upcoming month. We focus on predicting the direction of returns rather than their exact numerical values. This approach is particularly valuable in real-world scenarios where identifying the correct trend is more crucial for effective risk management than precisely estimating returns that may not align with market movements.

At each time point, the trained XGBoost model provides a probability $p_{i}$, indicating the likelihood that the next month's return for hedge fund $Y_{i}$ will be positive.

We then use $LTS_{i}$, $MRaR_{i}$, Sharpe Ratio, and $p_{i}$ as the primary features to evaluate hedge funds. Specific criteria based on these features are established to select hedge funds for investment. Only those funds predicted to perform strongly in the following month are retained in the portfolio. If a fund currently in the portfolio does not meet the selection criteria for the upcoming month, it is sold. Specifically, we set thresholds for each of the four features mentioned; a hedge fund must exceed all these thresholds to be considered likely to perform well in the next month. These thresholds are determined based on empirical analysis and historical market performance.

\section{Empirical Results}
In this section, we present a comprehensive overview of the data used for training and experimentation, outline the experimental design, and discuss the trading performance in detail.

\subsection{Data}\label{data}

In this study, we utilize a rich dataset containing monthly information on 10,545 hedge funds spanning from April 1994 to May 2023. As mentioned in the section on PolyModel theory, the data set is divided into two parts.

One part is the set of hedge funds which are also our target time series. Below is a sample of some of the representatives:

\indent

\begin{center}
\begin{tabular}{||c c ||} 
 \hline
 Label & Code \\ [1ex] 
 \hline\hline
\makecell{T-Bil} & {INGOVS USAB} \\
 \hline
 \makecell{SWAP 1Y Zone USA In\\USD DIRECT VAR-LOG} & {INMIDR USAB}  \\
 \hline
 \makecell{American Century Zero Coupon\\2020 Inv (BTTTX) 1989} & {BTTTX}  \\
 \hline
 \makecell{COMMODITY GOLD Zone USA\\In USD DIRECT VAR-LOG} & {COGOLD USAD} \\ 
 \hline
 \makecell{EQUITY MAIN Zone NORTH AMERICA\\In USD MEAN VAR-LOG} & {EQMAIN NAMM}  \\
\hline
 ...  & ...  \\
\hline
\end{tabular}
\captionof{table}{List of the Risk Factors for Hedge Funds Portfolio Construction}
\end{center}

\subsection{Benchmark Description}

\indent

The other part is the set of risk factors. We also list some of the representative ones below to illustarte how they look like:

\indent

\begin{center}
\begin{tabular}{||c c ||} 
 \hline
 Ticker & Name \\ [2.5ex] 
 \hline\hline
 U.S. Treasury  & LUATTRUU INDEX \\ 
 \hline
 U.S. Corp  & LUACTRUU INDEX  \\
 \hline
 Global High Yield  & LG30TRUU INDEX  \\
 \hline
 Heating Oil  & BCOMHO INDEX  \\
 \hline
 Orange Juice  & BCOMOJ INDEX  \\
 \hline
 Euro (BGN)  & EURUSD BGN Curncy  \\
\hline
Japanese Yen (BGN)  & USDJPY BGN Curncy  \\
\hline
 ...  & ...  \\
\hline

\end{tabular}
\captionof{table}{List of the Risk Factors for Network}
\end{center}

The PolyModel, described in section \ref{model}, is employed to derive 455 features in total from the raw data, which serve as the basis for our analysis.

However, due to the lack of mandatory reporting requirements by the Securities and Exchange Commission (SEC) for hedge funds, the dataset contains a significant number of missing values. To address this issue and enhance the quality of our analysis, we apply the following imputation methods for handling missing values:

\begin{itemize}
    \item If the values in `Return` are missing, we fill the missing value with -30.
    \item If the values in `Sharpe` are missing, we fill the missing value with -3.
    \item If the values in `LTS` are missing, we fill the missing value with -1.
    \item If the values in `MRaR` are missing, we fill the missing value with -3.
\end{itemize}

\subsection{Experimental Design}

As outlined in Section~\ref{feature}, the PolyModel generates 455 features from the raw data. We utilize the XGBoost algorithm~\cite{chen2016xgboost} to further analyze these features, specifically employing the XGBoost regressor to predict the returns of each hedge fund based on data from the previous month.

The XGBoost model is trained using a moving window approach, where two years of data serve as the training set, and predictions are made for the next month's returns across all 10,545 funds.

For comparison, we consider two benchmark trading strategies: the simple average and the weighted average. The simple average strategy allocates current cash evenly across all available hedge funds, while the weighted average strategy invests cash in hedge funds proportionally to their Assets Under Management (AUM). We assume monthly rebalancing of the portfolio, and transaction costs are not considered in this analysis.

This study focuses on three primary research questions: \begin{itemize} \item \textit{Does the use of machine learning methods improve performance compared to traditional non-machine learning methods?} \item \textit{How does the selection of features via the PolyModel influence trading performance?} \item \textit{Are larger funds consistently more reliable than smaller funds?} \end{itemize}

To address these questions, we design experiments considering several critical factors: the application of machine learning, the integration of PolyModel features, and the allocation of investments based on the AUM of the funds. Each experiment involves:

1) Deciding whether to use machine learning to select a subset of funds predicted to yield positive returns in the following month;
2) Determining whether to apply the PolyModel to filter funds based on predefined feature values such as LTS, Sharpe ratio, and MRaR;
3) Choosing whether to allocate funds based on the AUM of the funds.

Each experimental setup selects corresponding funds for investment and measures the resulting returns.

By exploring these questions, we aim to provide insights into the efficacy of machine learning techniques in forecasting hedge fund returns and the impact of feature selection on trading performance. Additionally, we examine the relationship between fund size and reliability, offering valuable perspectives for investment strategy formulation.

\subsection{Performance Evaluation}

\subsubsection{Evaluation Metrics}

To rigorously assess the performance of each trading strategy, we employ a comprehensive suite of evaluation metrics. The primary metrics include:

\begin{itemize}
    \item \textbf{Cumulative Return:} Measures the total return of the investment over the period of the study, providing a straightforward indicator of overall financial gain or loss.
    \item \textbf{Sharpe Ratio:} Evaluates the risk-adjusted return by comparing the return of the investment to its volatility, offering insights into the efficiency of the return relative to its risk.
    \item \textbf{Maximum Drawdown:} Captures the largest single drop from peak to trough in the investment value, highlighting the potential downside risk during the investment period.
\end{itemize}

In addition to these fundamental metrics, we calculate a range of supplementary measures to provide a more detailed characterization of the trading strategies. These additional metrics include:

\begin{itemize}
    \item \textbf{Annual Return:} The yearly rate of return, which helps in comparing the performance across different time scales and market conditions.
    \item \textbf{Annual Volatility:} Measures the standard deviation of the investment's returns on an annual basis, indicating the level of risk or fluctuation in returns over time.
    \item \textbf{Count of Months with Positive Returns:} Tracks the number of months during which the strategy yielded a profit, reflecting consistency and reliability.
    \item \textbf{Count of Months with Negative Returns:} Counts the months with losses, providing a counterbalance to the positive return metric and highlighting potential volatility or risk.
    \item \textbf{Maximum Monthly Increase:} Identifies the highest single-month gain, offering a peak into the potential for rapid growth.
    \item \textbf{Maximum Monthly Decrease:} Indicates the most significant single-month loss, which is crucial for understanding the worst-case scenarios.
    \item \textbf{Average Monthly Increase:} Calculates the average of all monthly gains, providing a measure of typical positive performance.
    \item \textbf{Correlations with Established Hedge Fund Indices:} We also examine the correlations with major hedge fund indices such as the HFRIFOF (Hedge Fund Research Investable Fund of Funds Index) and HFRIFWI (Hedge Fund Research Investable World Index), which helps in understanding how closely the strategy aligns with broader market movements.
\end{itemize}

By integrating this diverse array of evaluation metrics, we ensure a thorough and nuanced analysis of the trading strategies. This multifaceted approach not only enhances our understanding of each strategy's performance but also aids in making informed comparisons, thereby identifying the strengths and weaknesses more effectively.

\subsubsection{Impact of Machine Learning on Trading Performance}

Table~\ref{tab:ml} presents a comparative analysis of average trading performance between experiments that utilize machine learning for fund selection and those that do not. The data decisively address the question of whether machine learning enhances trading performance. Although the Sharpe ratio is marginally higher in scenarios without machine learning-based fund selection, the average cumulative return is notably greater when machine learning is employed.

This enhancement in cumulative return suggests that machine learning-based methods lead to a more strategically focused portfolio. Notably, the non-machine learning approach tends to select a broader array of funds, potentially diluting overall performance but decreasing portfolio volatility due to increased diversification.

Moreover, while the frequency of positive returns is greater in the non-machine learning experiments, those utilizing machine learning demonstrate superior outcomes in terms of maximum monthly increase and average monthly increase. This indicates that machine learning is particularly effective at pinpointing the most promising funds within the available pool. However, it also underscores the inherent variability in machine learning performance, which can result in higher annual volatility in portfolios that integrate these techniques.

In conclusion, the findings underscore that machine learning can significantly enhance cumulative returns. However, this advantage comes at the cost of increased annual volatility, reflecting the fluctuating nature of machine learning-based predictions.

\begin{table}[htbp]
  \centering
  \caption{Comparison of trading performance between using and without using machine learning }
\begin{tabular}{lrr}
\hline
Using Machine Learning &       False &        True \\
\hline
Cumulative returns        &   19.592471 &   24.053971 \\
Sharpe Ratio              &    1.200638 &    1.181875 \\
Max Drawdown              &    0.237923 &    0.245882 \\
Number of Months Increase &  219.285714 &  217.428571 \\
Number of Months Decrease &   89.785714 &   89.142857 \\
Max Monthly Increase      &    0.156736 &    0.174337 \\
Max Monthly Decrease      &   -0.186923 &   -0.181905 \\
Average Monthly Increase  &    0.021822 &    0.023683 \\
Annual Return             &    0.116122 &    0.126893 \\
Annual Volatility         &    0.098003 &    0.105883 \\
Correlation with HFRIFOF  &    0.134775 &    0.109414 \\
Correlation with HFRIFWI  &    0.155489 &    0.123860 \\
\hline
\end{tabular}
\label{tab:ml}%
\end{table}

\subsubsection{Influence of PolyModel Feature Selection}

Table~\ref{tab:filter} provides an overview of the average trading performance across experiments that utilize various combinations of filters for fund selection. These filters are based on predefined criteria for LTS, Morningstar, MRaR, and Sharpe Ratio. This analysis seeks to elucidate the impact of PolyModel feature selection on trading performance.

The data from these experiments unequivocally show that feature selection significantly influences trading outcomes. Strategies that incorporate all available features for fund selection consistently outperform those that employ a more limited set of features or no feature-based filtering at all. Moreover, the results indicate that the use of a comprehensive set of filters not only enhances cumulative returns but also contributes to a more robust trading strategy.

A particularly notable finding from this study is the role of LTS as a standalone filter. When LTS is employed as the primary feature for fund selection, the experiments yield an average cumulative return of 10.24, coupled with an impressive average Sharpe ratio of 1.65. This highlights LTS's pivotal role in reducing portfolio volatility while maintaining favorable returns.

These findings underscore the critical importance of strategic feature selection in optimizing hedge fund portfolios. By effectively leveraging PolyModel features, investors can significantly enhance both the performance and stability of their investment strategies.

\begin{table}[htbp]
  \centering
  \caption{Comparison of trading performance among combinations of filter usage}
\begin{tabular}{lrrr}
\hline
{} &  cumulative returns &  Sharpe Ratio &  Max Drawdown \\
Filters         &                     &               &               \\
\hline
No use                &            9.336700 &      1.301033 &      0.172686 \\
LTS             &           10.244457 &      1.650248 &      0.088949 \\
LTS, MRaR        &           28.498878 &      1.179584 &      0.337729 \\
LTS, Sharpe      &           23.634776 &      1.118033 &      0.266081 \\
LTS, Sharpe, MRaR &           30.590464 &      1.184490 &      0.337729 \\
MRaR            &           19.365154 &      1.082842 &      0.217072 \\
Sharpe          &           21.180221 &      1.060660 &      0.220558 \\
Sharpe, MRaR     &           19.248598 &      1.062941 &      0.225198 \\
\hline
\end{tabular}
\label{tab:filter}%
\end{table}

\subsubsection{Reliability of Larger Funds}

Table~\ref{tab:weight} presents a comparative analysis of average trading performance among experiments that employ different combinations of filters for fund selection. This section specifically addresses the question of whether larger funds, as indicated by their Assets Under Management (AUM), are consistently more reliable than smaller funds.

Contrary to common expectations, the results reveal that strategies allocating money proportionally based on funds' AUM do not outperform strategies that distribute money evenly across all funds. This observation is supported by all key performance indicators used in the study. Such findings challenge the prevailing assumption that larger funds are inherently more reliable and profitable than their smaller counterparts.

These results suggest that fund size, as measured by AUM, may not be a reliable indicator of fund performance. This insight calls for a reevaluation of investment strategies that prioritize larger funds based solely on their size, advocating for a more nuanced approach that considers a broader range of performance metrics and fund characteristics.

\begin{table}[htbp]
  \centering
  \caption{Comparison of trading performance between allocating money evenly or weighted by funds' AUM}
\begin{tabular}{lrr}
\hline
Weighted &       False &        True \\
\hline
Cumulative returns        &   28.090621 &   15.555821 \\
Sharpe Ratio              &    1.339130 &    1.043383 \\
Max Drawdown              &    0.237112 &    0.246692 \\
Number of Months Increase &  223.928571 &  212.785714 \\
Number of Months Decrease &   84.214286 &   94.714286 \\
Max Monthly Increase      &    0.143105 &    0.187968 \\
Max Monthly Decrease      &   -0.195470 &   -0.173357 \\
Average Monthly Increase  &    0.022723 &    0.022782 \\
Annual Return             &    0.133733 &    0.109282 \\
Annual Volatility         &    0.098141 &    0.105744 \\
Correlation with HFRIFOF  &    0.124408 &    0.119781 \\
Correlation with HFRIFWI  &    0.136393 &    0.142956 \\
\hline
\end{tabular}
\label{tab:weight}%
\end{table}

\subsubsection{Performance Showcase of the Best Performer}

Table~\ref{tab:best} highlights the best performer selected from the experimental results. This top-performing strategy is distinguished by its comprehensive use of all three predefined features as filters, the application of machine learning for fund selection, and an even distribution of funds across selected investments.

\begin{figure}[htbp]
\centerline{\includegraphics[width=10cm]{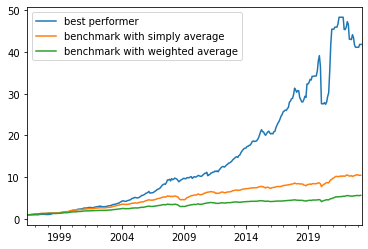}}
\caption{Best Performer trading performance}
\label{best_pic}
\end{figure}

Figure~\ref{best_pic} provides a visual representation of the cumulative return plot for the best performer, juxtaposed against two benchmark strategies. This graphical illustration clearly demonstrates the significant enhancement in trading performance that can be achieved through a synergistic approach combining machine learning-based fund selection with PolyModel feature selection.

The success of the best performer not only validates the effectiveness of integrating advanced analytical techniques but also serves as a powerful example of how strategic fund selection and asset allocation can lead to superior investment outcomes. This case exemplifies the potential benefits of adopting sophisticated, data-driven strategies in hedge fund management.

\begin{table}[htbp]
  \centering
  \caption{Best Performer}
\begin{tabular}{ll}
\hline
{} &               25 \\
\hline
Filters                   &  LTS,Sharpe,MRaR \\
Using Machine Leaerning                    &             True \\
Weighted                  &            False \\
Cumulative returns        &        41.814637 \\
Number of Months Increase &              221 \\
Number of Months Decrease &               74 \\
Max Monthly Increase      &           0.1789 \\
Max Monthly Decrease      &        -0.243933 \\
Average Monthly Increase  &         0.024882 \\
Annual Return             &         0.154407 \\
Annual Volatility         &         0.108745 \\
Sharpe Ratio              &         1.341802 \\
Max Drawdown              &         0.298017 \\
Correlation with HFRIFOF  &         0.128749 \\
Correlation with HFRIFWI  &          0.13045 \\
\hline
\end{tabular}
\label{tab:best}%
\end{table}

\section{Conclusion}

In this study, we conducted a comprehensive examination of the impact of machine learning methods, PolyModel feature selection, and fund size on hedge fund investment strategies. Our investigation aimed to illuminate several critical questions of substantial interest to both researchers and practitioners in the field of financial investments.

Our findings have provided valuable insights into the key research questions addressed in this study:

\begin{itemize}
    \item \textbf{Machine Learning in Hedge Fund Strategies:} We explored the performance of hedge fund investment strategies with and without the utilization of machine learning methods. The results revealed that, although the Sharpe ratio may exhibit marginal improvements without machine learning, the average cumulative return significantly benefits from the use of machine learning. This suggests that machine learning can effectively identify a subset of funds with the potential for positive returns, ultimately leading to enhanced overall performance.
    
    \item \textbf{Impact of PolyModel Feature Selection:} We delved into the influence of PolyModel feature selection on trading performance. Our analysis indicated that employing all available features for fund selection consistently outperforms experiments that use fewer features or no feature-based filtering. Furthermore, the use of multiple filters correlated with PolyModel features was associated with higher cumulative returns. Notably, Long-Term Stability (LTS) emerged as a key feature capable of controlling portfolio volatility, while still generating favorable returns.
    
    \item \textbf{Reliability of Larger Funds:} We examined the common assumption that larger funds are invariably more reliable than smaller ones. Our results challenged this conventional wisdom, as experiments allocating money proportionally to fund AUM did not surpass those evenly distributing funds. This counterintuitive finding suggests that fund size does not consistently translate to increased reliability in investment outcomes.
\end{itemize}

In conclusion, this research underscores the transformative potential of machine learning and PolyModel feature selection in enhancing hedge fund investment strategies. While these methods introduce a degree of variability and challenges, they also offer promising avenues for improving cumulative returns. Following a similar spirit, \cite{zhao2024hedgefundportfolioconstruction} considers the combination of PolyModel feature construction with deep learning - iTransformer for hedge fund portfolio construction, and also achieve a very good cumulative return. Moreover, the size of a fund should not be equated with its reliability, and more nuanced considerations are required to design effective investment strategies. The insights garnered from this study hold practical significance for investors and asset managers seeking to optimize their hedge fund portfolios, contributing to the evolving landscape of data-driven financial decision-making.

\newpage
\bibliographystyle{chicago}
\bibliography{ref}
\end{document}